\documentclass[pre,aps,twocolumn,showpacs,floats]{revtex4}

\usepackage{graphics,graphicx,epsfig}
\usepackage{bm}
\begin{document} 

\title{Tricriticality of the Blume-Emery-Griffiths Model in Thin Films of Stacked Triangular Lattices}
\author{Sahbi EL HOG and H. T. DIEP}
\affiliation{Laboratoire de Physique Th\'eorique et Mod\'elisation \\
Universit\'e de Cergy-Pontoise, CNRS, UMR 8089 \\ 2, avenue Adolphe Chauvin,\\
Cergy-Pontoise Cedex, France}
\date{\today}

\begin{abstract}
We study in this paper the Blume-Emery-Griffiths model in a thin film of stacked triangular lattices.
The model is described by three parameters: bilinear exchange interaction between spins $J$, quadratic exchange  interaction $K$ and single-ion anisotropy $D$. The spin $S_i$ at the lattice site $i$ takes three values $(-1,0,+1)$.
This model can describe the mixing phase of He-4 ($S_i =+1,-1$) and He-3 ($S_i =0$) at low temperatures.
Using Monte Carlo simulations,  we show that there exists a critical value of $D$ below (above) which the transition is of second-(first-)order.
In general, the temperature dependence of the concentrations of He-3 is different from layer by layer. At a finite temperature in the superfluid phase, the film surface shows a deficit of He-4 with respect to interior layers. However, effects of surface interaction parameters can reverse this situation. Effects of the film thickness on physical properties will be also shown as functions of temperature.
\end{abstract}


\maketitle
\section{Introduction}
The physics of thin films has seen a spectacular development during the last 30 years. This is due, on the on hand, to numerous electronic applications using thin films  \cite{Heinrich,Zangwill,DiepTM}, superlattices and multilayers \cite{Grahn}, and on the other hand, to the lack of theoretical understanding of surface properties which are very different from the bulk ones.  Since it is rather easy to change the conditions at the surface of a films, by coating or by adsorption of other species, for instance,  surface physics offers a
lot of opportunities to discover new microscopic phenomena leading to potential electronic applications. One has seen in recent years applications using phenomena such as giant magnetoresistance \cite{Fert,Grunberg}, spin transfer torques \cite{Stiles}, spin valves, etc.\cite{Dietl}.

Theoretically, surface effects in thin films such as surface phonons, surface magnons, surface plasmons have been widely studied.  We will focus in this paper on the magnetic properties of thin films.  In surface magnetism, much has been understood, in particular on the existence of surface-localized spin-waves and their effects on physical behaviors of thin films at finite temperatures such as the reduction of the critical temperature and the low surface magnetization \cite{Diep1979,NgoSurface,Diep2015}. In most of previous studies, the spin models such as Ising and Heisenberg models have been widely investigated.

In this paper, we use the  Blume-Emery-Griffiths (BEG) model to study physical behaviors of thin films. The spin in this model has three states +1, -1 and 0.  A site with a value 0 represents a vacant site. The system is considered as a dilute magnetic systems in which the number of vacant sites varies as a function of temperature ($T$).
This model can also describe the mixing phase of superfluid He-4 ($S_i =+1,-1$) and normal fluid He-3 ($S_i =0$) at low temperatures \cite{Puha,BEG}.  Other models extended from the original BEG model have been recently introduced to study the effects of vacancies and of the continuous degrees of freedom in the mixtures He-3 and He-4 \cite{dietrich1,dietrich2}.

The present work has been motivated by the desire, on the one hand, to know if results of bulk BEG model remain valid in films, and on the other hand to see if the transition criticality can be altered when we reduce the film thickness as we have seen in Ref. \onlinecite{Pham1} and \onlinecite{Pham2}: (i) there is a cross-over from 3-dimensional (3D) criticality to 2-dimensional (2D) universality with decreasing thickness for a second-order transition \cite{Pham1}, and (ii) the 3D first-order transition becomes a second-order transition at very small thickness \cite{Pham2}.  We note that the BEG model has been studied by a number of authors in thin films of simple cubic lattice structure \cite{Balcerzak1,Balcerzak2,Ez-Zahraouy,Jabar} but the motivation was not the same as the present paper. In Refs. \onlinecite{Balcerzak1} and \onlinecite{Balcerzak2} the authors have used the mean-field approximation to study the phase diagram of a double-layer and 5-layer films using negative values of biquadratic $K$ term in Eq. (\ref{BEG1}) below. They found a very rich phase diagram with a tricritical point and a staggered quadrupolar phase. The authors of Ref. \onlinecite{Ez-Zahraouy} studied the same model of 5-layer film with negative $K$ by mean-field theory. To our knowledge, the mean-field theory cannot be used to determine precise phase diagrams and the criticality in general in 2D and 3D systems. In Ref. \onlinecite{Jabar}, the authors have investigated a system of mixed spins $S=3/2$ and 2 using the BEG model. They determined the phase diagram in various parts of the phase space. Very simple Monte Carlo (MC) simulations have been carried out.  In particular, too few data were obtained in the critical phase transition region to be useful for the determination of the transition temperature.

Section II is devoted to the description of the model and the simulation method. Results will be shown and discussed in Sect. III. Concluding remarks are given in Sect. IV.

\section{Model and method}
\subsection{Model}
The Blume-Emery-Griffiths (BEG) model consists of a system with three states per spin. The model is described by the Hamiltonian
\begin{equation}\label{BEG1}
 H=-J\sum\limits_{<i,j>} S_iS_j-K\sum\limits_{<i,j>}S_i^2S_j^2+D\sum\limits_{i}S_i^2
\end{equation}
where the spin variable takes the value $S_i=-1,0,1$ and $\sum_{<i,j>}$ denotes a summation over all nearest-neighbors (NN).
The model presents in addition to the bilinear spin interaction $J$ a biquadratic interaction $K$ and a single-ion crystal field $D$.
This model can describe the mixing phase of He-4 ($S_i =+1,-1$) and He-3 ($S_i =0$) at low temperatures \cite{Puha,BEG}.

The present paper studies this model in a film composed of $L_z$ infinite $xy$ triangular lattices stacked in the $z$ direction. 
We have chosen the stacked triangular lattices to have a larger coordination number of neighbors. Since we worked with very thin films of small quantities of matter, such a large coordination number reduces numerical errors on statistical fluctuations.
Each site is occupied by a spin of values $\pm 1,0$. In the Helium language, these spins are atoms He-3 and He-4. Since we are working at a given finite temperature $T$ (canonical method) we leave the system to determine the concentration of He-3 and He-4 at equilibrium at each given $T$. The mixing of He-3 and He-4 is at random corresponding to the maximum entropy. As will be seen below, when there is a surface the He-3 concentration is more important near the surface than in the interior.

In the bulk, if $D=0$ then the model is Ising-like. Nonzero values of chemical potential $D$ favor the proliferation of zero spins in the system and lower the transition temperature between the superfluid and the normal fluid phase. By increasing $D$ the system can support superfluid ordering but with a mixture of the normal liquid (He-3) and the superfluid one (He-4). The interplay between the superfluid-like ordering bilinear term and the phase breaking $D$ term generates an exotic phase diagram that consists of a line of continuous phase transitions at low $D$ values and high temperatures \cite{BEG,dietrich1,dietrich2}. At high $D$ and low $T$ values, the transition becomes discontinuous.

We will show below that the bulk feature of the phase diagram in the space ($T,D)$ is found for thin films though there is a variation of the critical value of $D$ above which the transition is of first order (see next section).  We will also show that the phase diagram depends on the surface parameters.

\subsection{Method}
We use MC simulation \cite{Binder} to calculate properties of the system at finite temperatures for the  size of $L\times L\times L_z$ where $L_z$ is the film thickness.  Periodic boundary conditions are used in the $xy$ plane. The standard MC  method is used to study the phase transition. The averaged energy
and the specific heat per spin are defined by
\begin{equation}
 \langle E\rangle=\frac{\langle H\rangle}{N}
\end{equation}
\begin{equation}
 C_V=N\frac{\langle E^2\rangle-\langle E\rangle^2}{k_BT^2}
\end{equation}
where $\langle...\rangle$ indicates the thermal average. We define the order parameter $Q$ for the $q$-state Potts model by
\begin{equation}
 Q=[q\max(Q_1,Q_2,Q_3)-1]/(q-1)
 \end{equation}
 where $Q_n$ is the spatial average defined by
 \begin{equation}
  Q_n=\frac{1}{N}\sum\limits_{i=1}^N\delta (S_i,n)
 \end{equation}
 $n=-1,+1,0$ indicates  the value of spin $S_i$ at site $i$, $\delta (S_i,n)$ the Kronecker symbol, and $N$ the total number of sites. 
 Our choice of the 3-state Potts order parameter is motivated by the fact that
the values of the parameter -1, +1 and 0 make it impossible to use the usual definition of the magnetization where one sums all values of spins. It is however to be used with complementary quantities such as the average number of each of the values 1, -1 and 0 because the Potts order parameter can tell us if there is an ordering but it does not give the information on the value of spin which makes the system ordered. 
The susceptibility per spin is defined by
 \begin{equation}
 \chi=N\frac{\langle Q^2\rangle-\langle Q\rangle^2}{k_BT}
\end{equation}

In general, we discard about $10^5$ MC steps per spin to equilibrate the system at temperature $T$ before averaging physical quantities over the next $10^5$ MC steps. 
We shall show below an example of the time evolution of the energy and the order parameter. 
For histograms, we recorded in general  $10^6$ MC steps per spin. The lattice sizes used in our simulations are $L=20,30,...,120,300$ and $L_z=4,8,12,16$.

\section{Results}
\begin{figure}
\centering
\includegraphics[width=6cm]{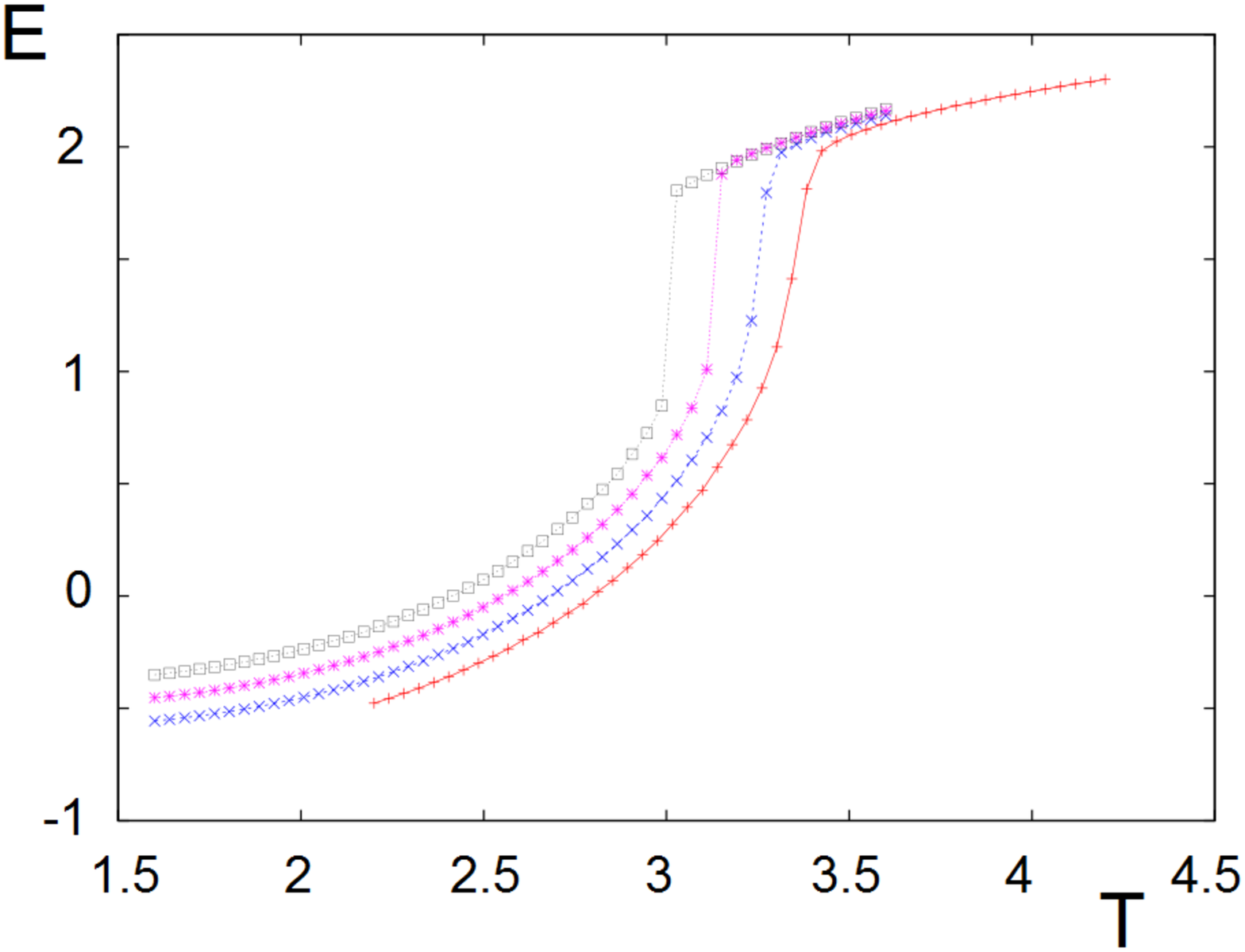}
\includegraphics[width=6cm]{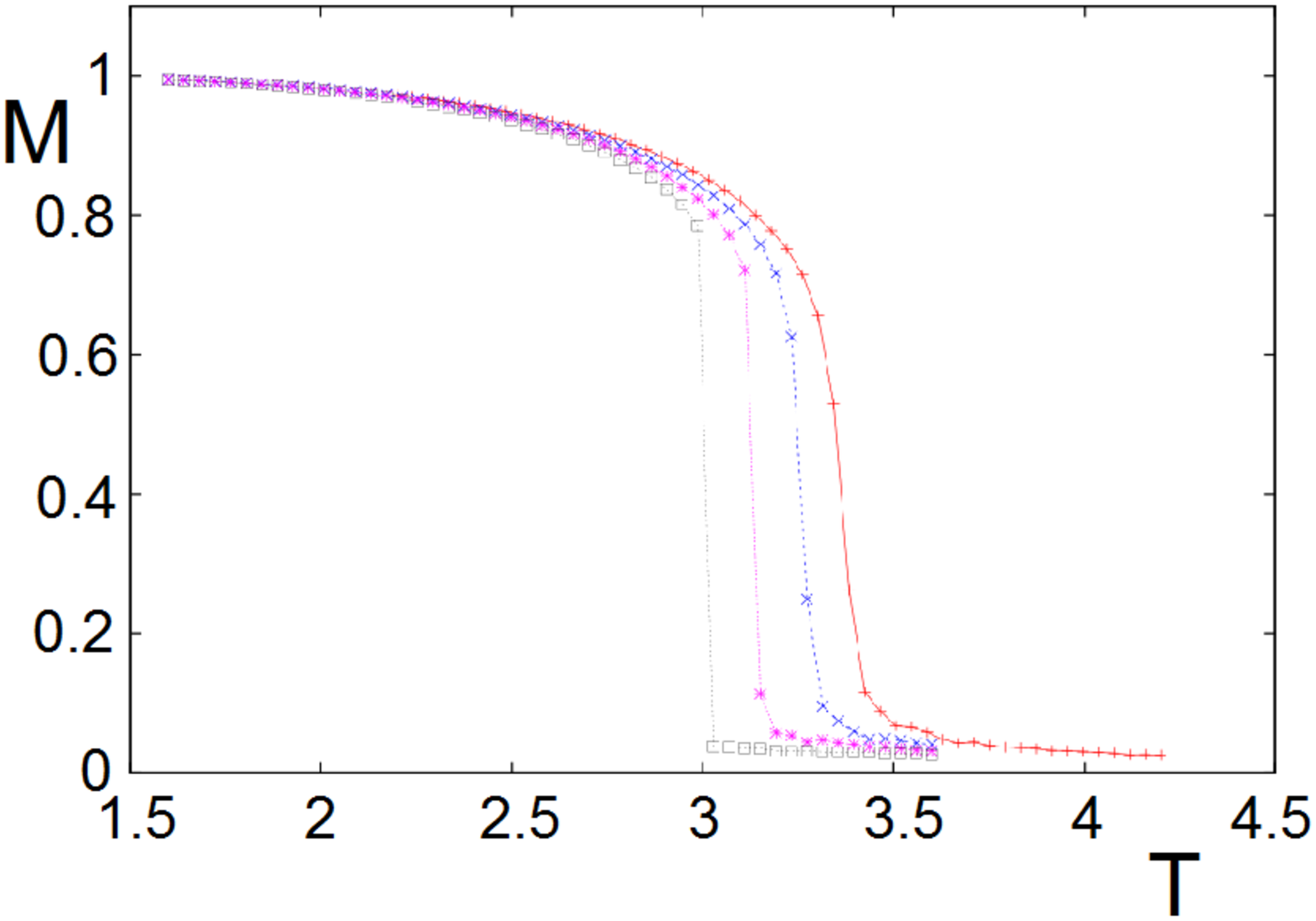}
\includegraphics[width=6cm]{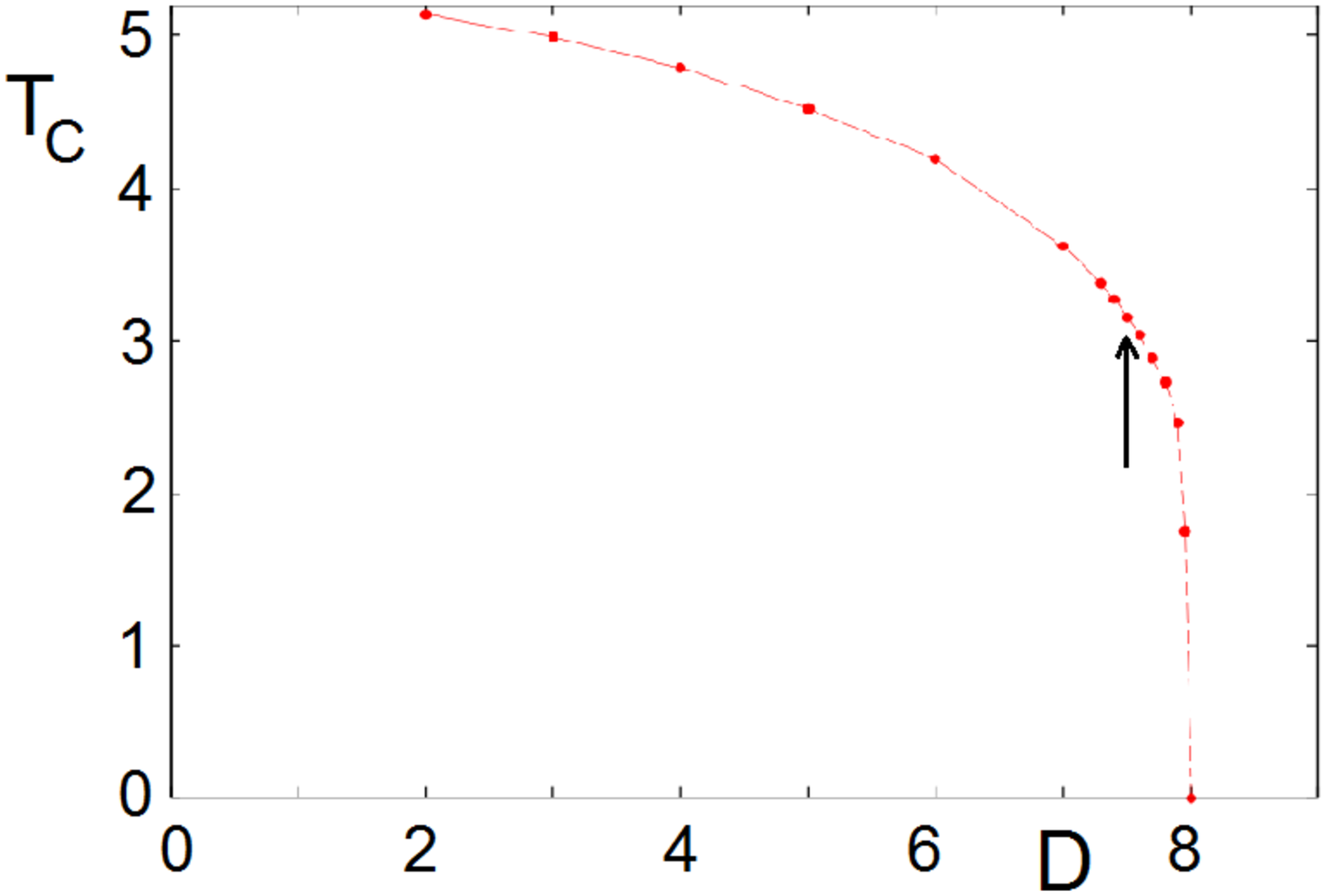}
\caption{\label{bulk}  (Color online) Bulk properties: Energy (top) and magnetization (middle) versus $T$ around the tricritical $D$ of the bulk case. From right to left: $D=7.3$, 7.4, 7.5 and 7.6. The cross-over from the second order to the first order occurs at $D_c\simeq 7.5$. The transition temperature versus $D$ is shown in the bottom figure: the arrow indicates the bulk tricritical point. Simulations have been done for a crystal of $20^3$ sites with periodic boundary conditions in all directions.}
\end{figure}

Let us take $J=1$ and $K=1$, namely ferromagnetic interactions between NN. 
Before showing the results for thin films, let us show first the results for the bulk properties of of the BEG model applied to the stacked triangular lattices. These results by symmetry argument do not bring new physics with respect to the case of simple cubic lattice \cite{BEG}. However, these results provide elements for comparison with the film case which will be shown in details below.
We show in Fig. \ref{bulk} the bulk case: the top figure shows the energy $E$ versus temperature $T$ for several values of $D$ in the tricritical region. As seen, the transition is continuous for $D<D_c\simeq 7.5$ and discontinuous for $D>D_c$.
The magnetization (middle figure) shows also a discontinuity for $D>D_c$. The bottom figure shows the transition curve in the space $(D,T)$ where the tricritical point is indicated by an arrow. Technical details on the determination of $D_c$ are similar to those used in the case of thin films. They will be given below.

For a given film thickness, we study in the same manner the behavior of the BEG model for different values
of $D$ by calculating  the energy, specific heat, the order parameter, the layer magnetization and the energy histogram.

\subsection{Order of the phase transition}

\begin{figure}
\centering
\includegraphics[width=6cm]{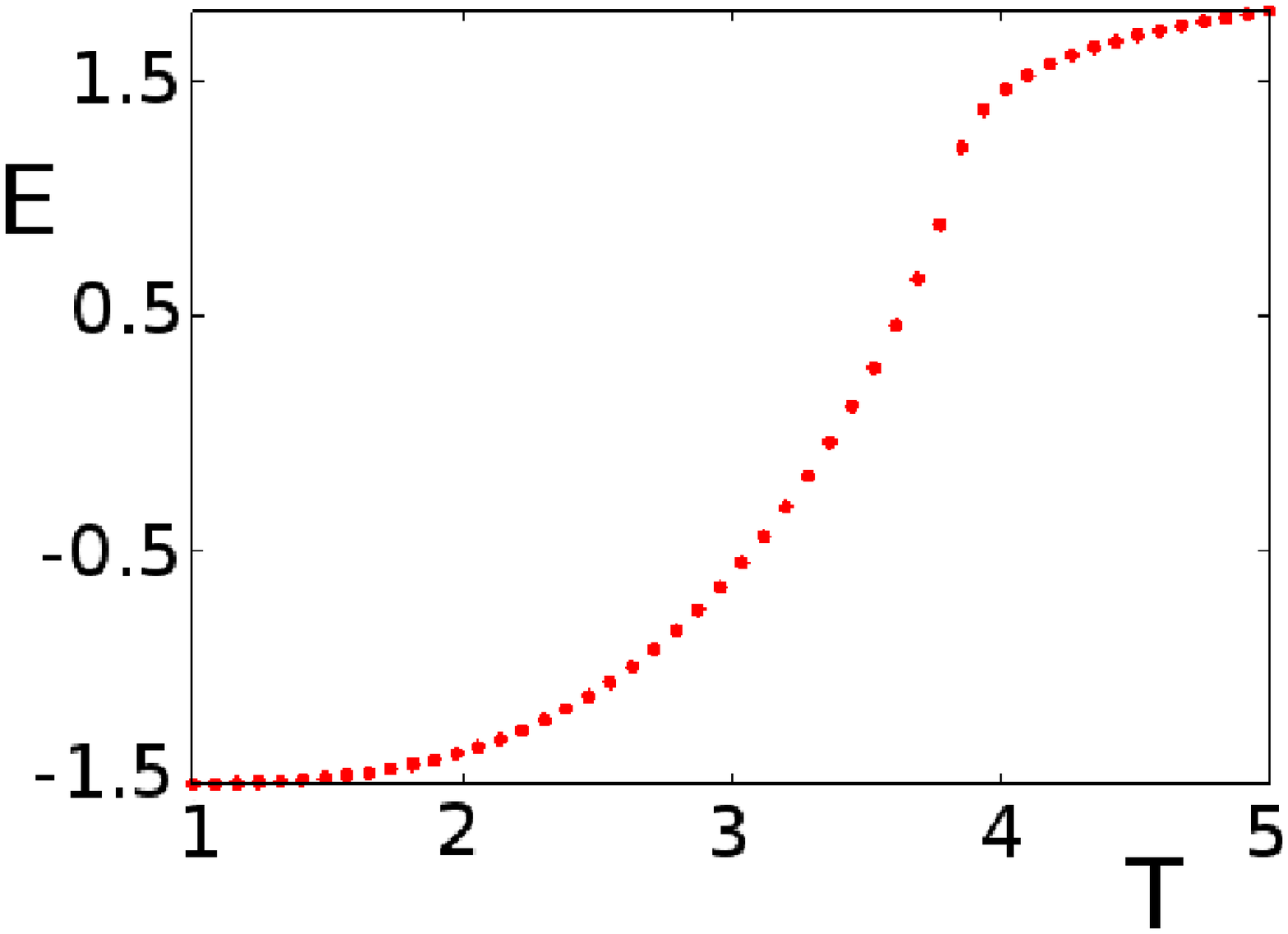}
\includegraphics[width=6cm]{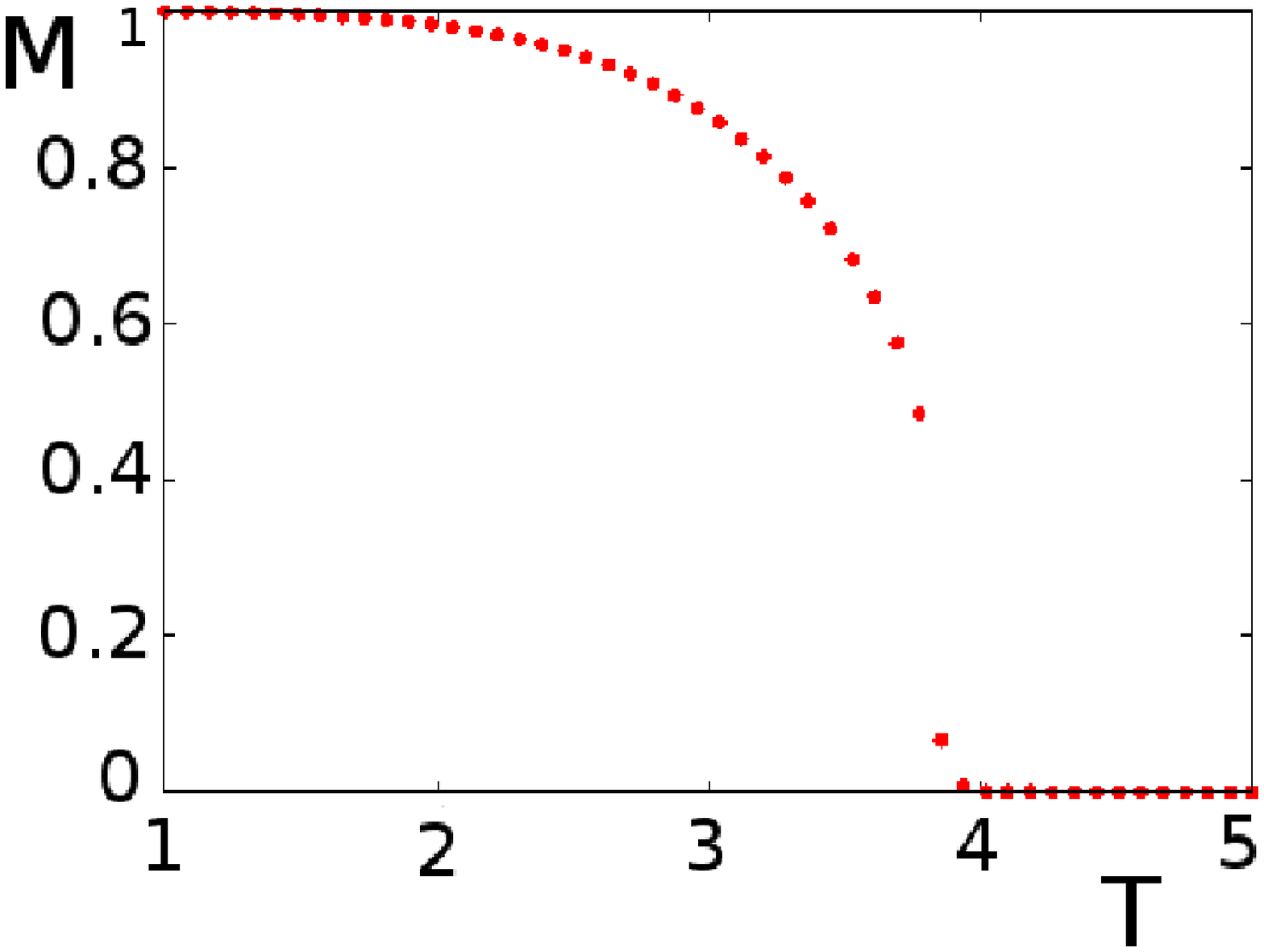}
\includegraphics[width=6cm]{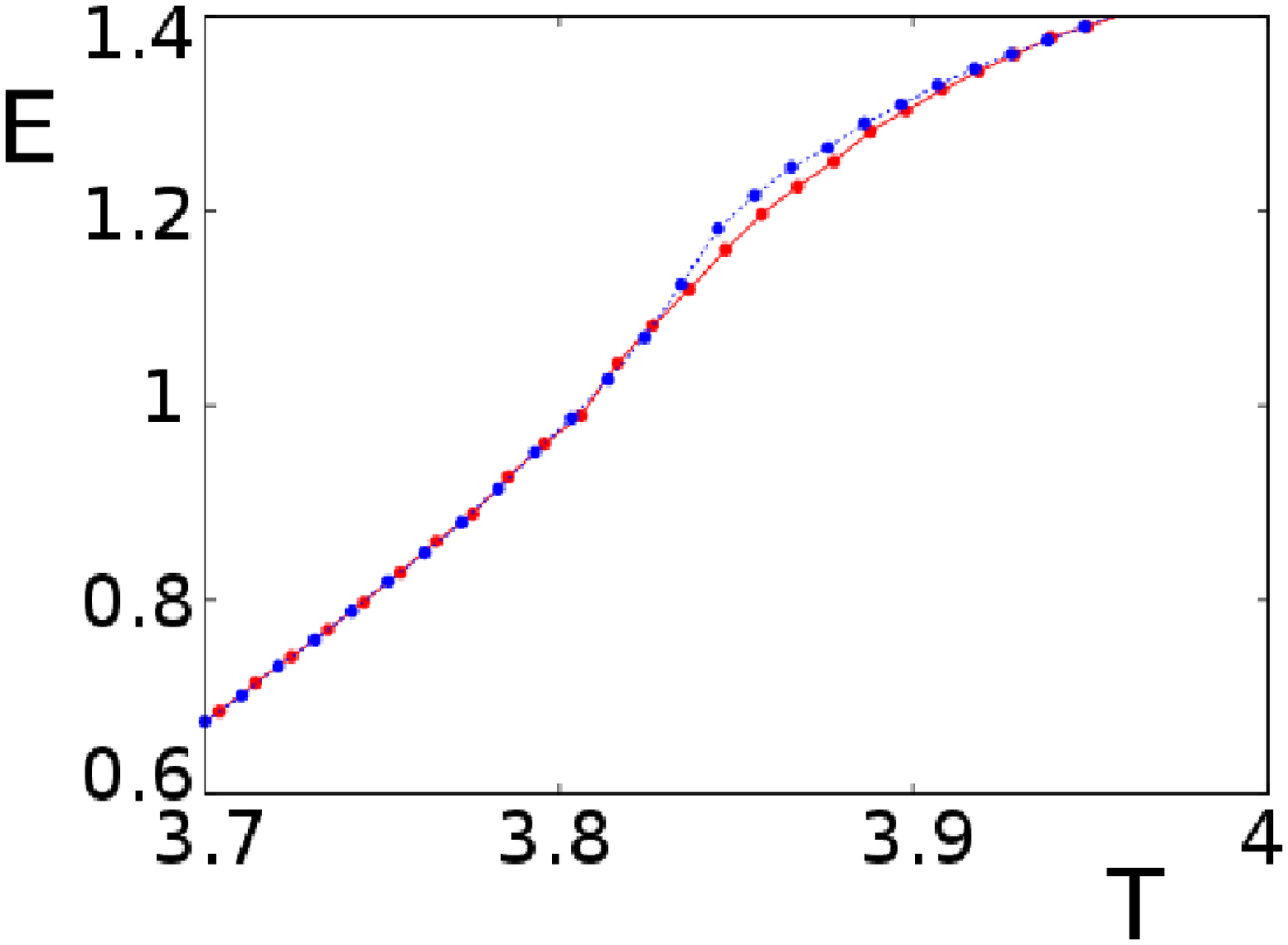}
\caption{\label{fed6}  (Color online) Energy $E$ (top) and magnetization $M$ (middle) versus $T$ for $D=6$, $L_z=4$ and $L=120$.
The size effect on $E$ in the transition region is zoomed for $L=36$ (red) and 300 (blue). The size effect on $M$ for those sizes is not clearly distinguished.}
\end{figure}

In Fig. \ref{fed6} we show the energy $E$ and magnetization $M$ versus $T$ in the case where $D=6$, $L_z=4$ with $L=120$. Note that the lateral size effect is slightly seen in $E$ as zoomed in the bottom figure, but it is not distinguishable in $M$.
The curves $E$ and $M$ present a second-order
phase transition at $T_c\simeq 3.82$.

\begin{figure}
\centering
\includegraphics[width=6cm]{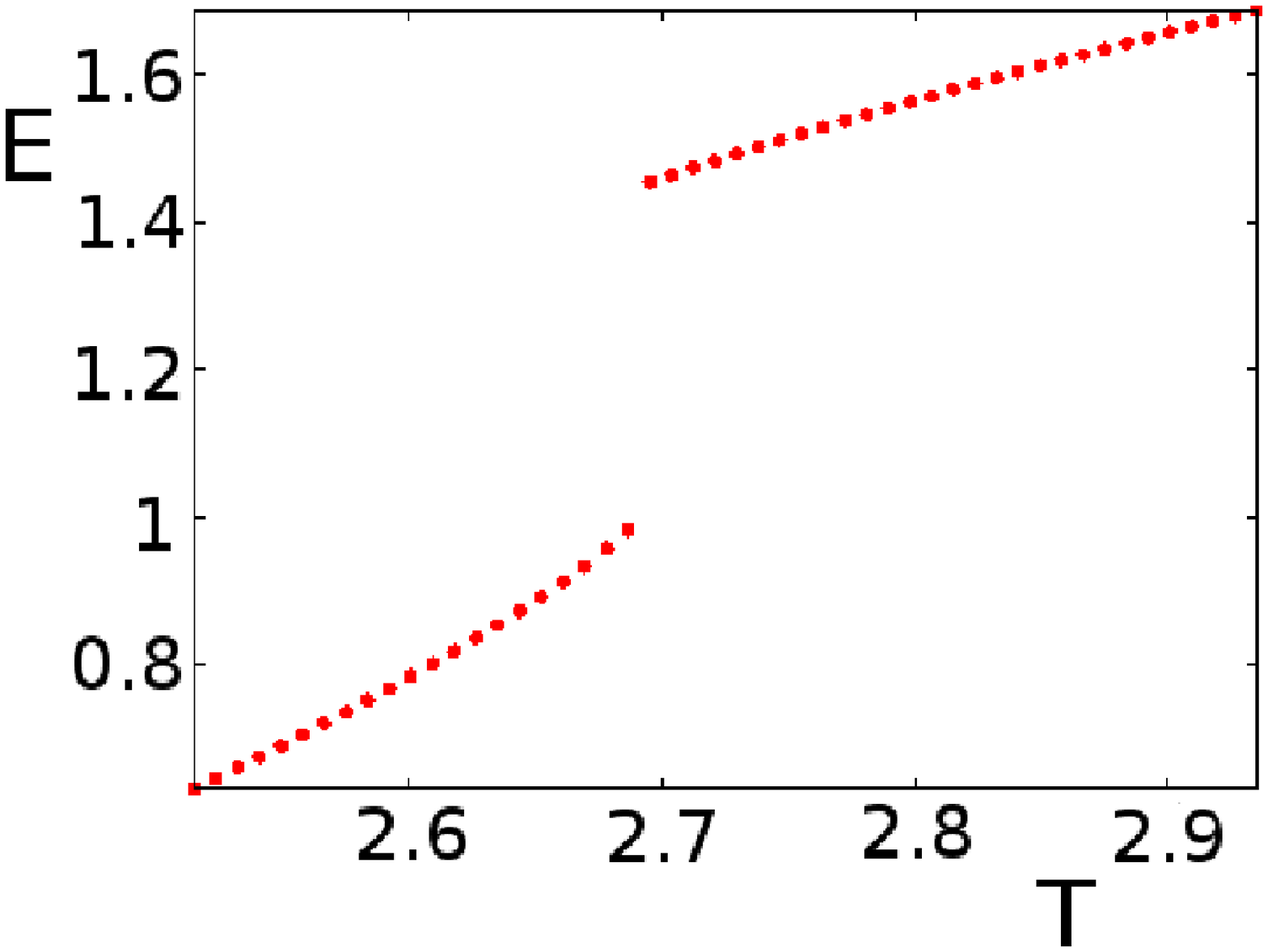}
\includegraphics[width=6cm]{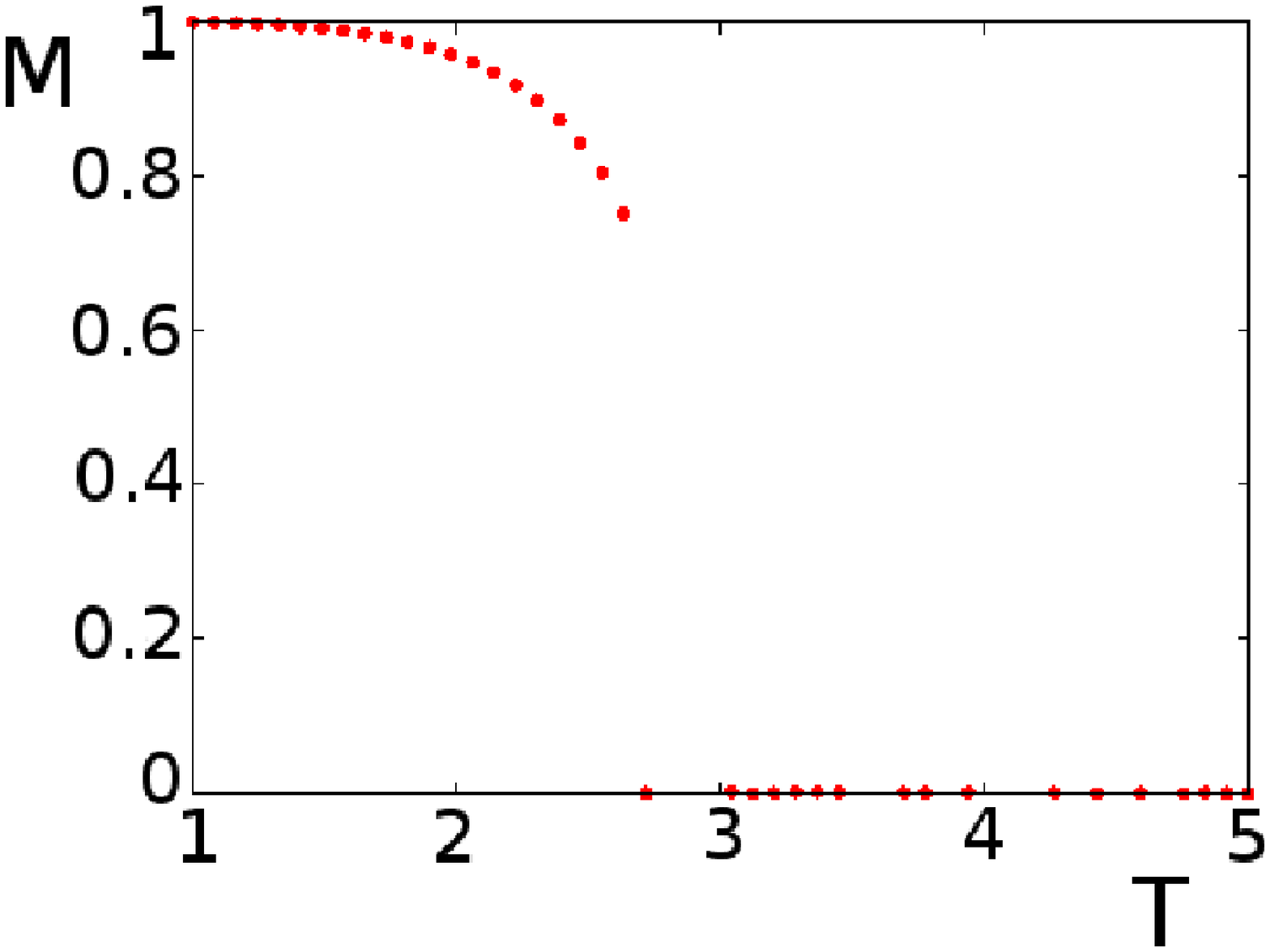}
\caption{\label{fed73}  (Color online) Energy (top) and magnetization (bottom) versus $T$ in the first-order region of $D$: $D=7.3$, $L=120$, $L_z=4$.}
\end{figure}

With increasing $D$, the system undergoes a first-order transition. We show in Fig. \ref{fed73} the case of $D=7.3$ where one observes a discontinuity at the transition temperature $T_c\simeq 2.694$.

\begin{figure}
\centering
\includegraphics[width=6cm]{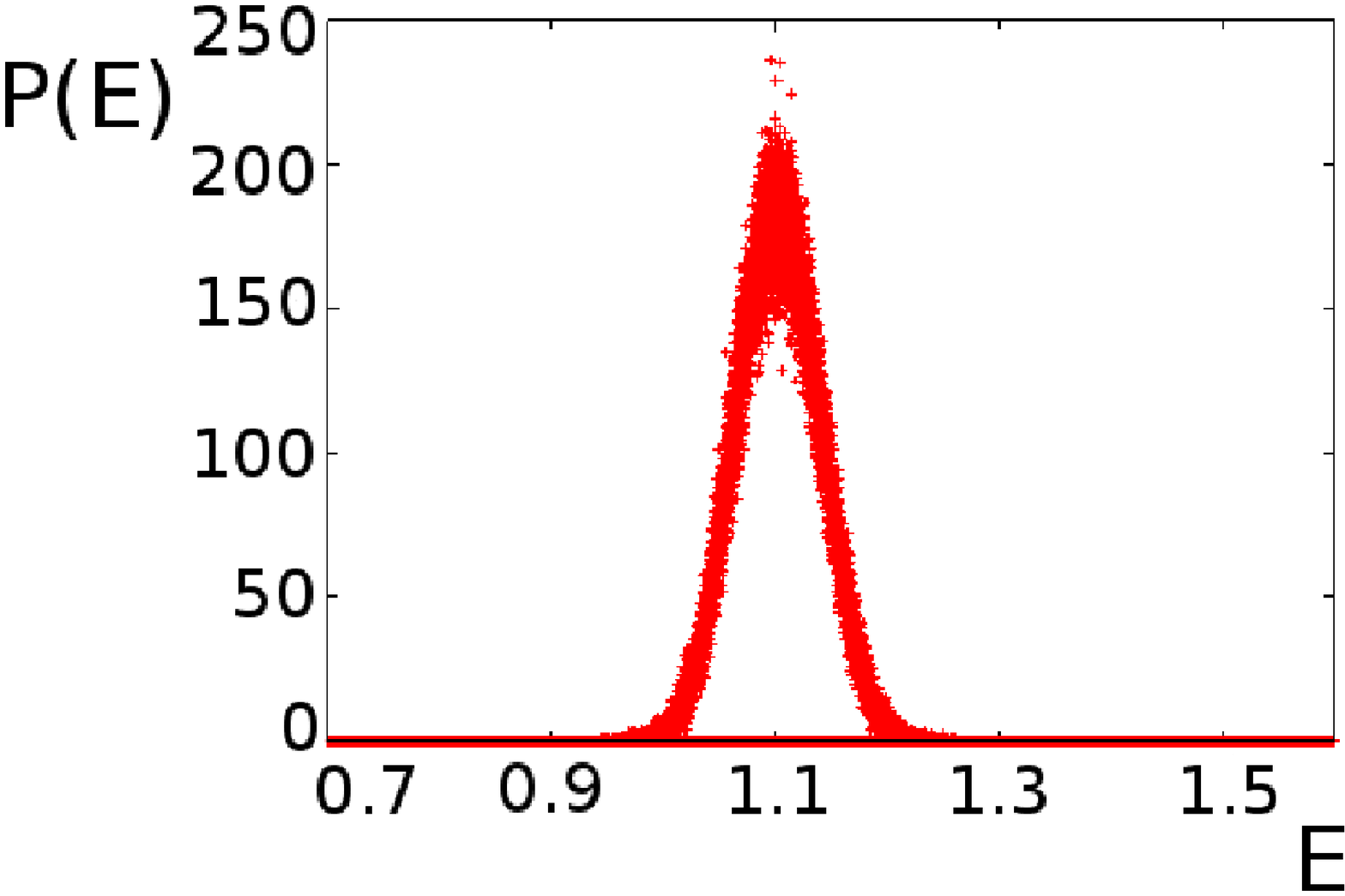}
\includegraphics[width=6cm]{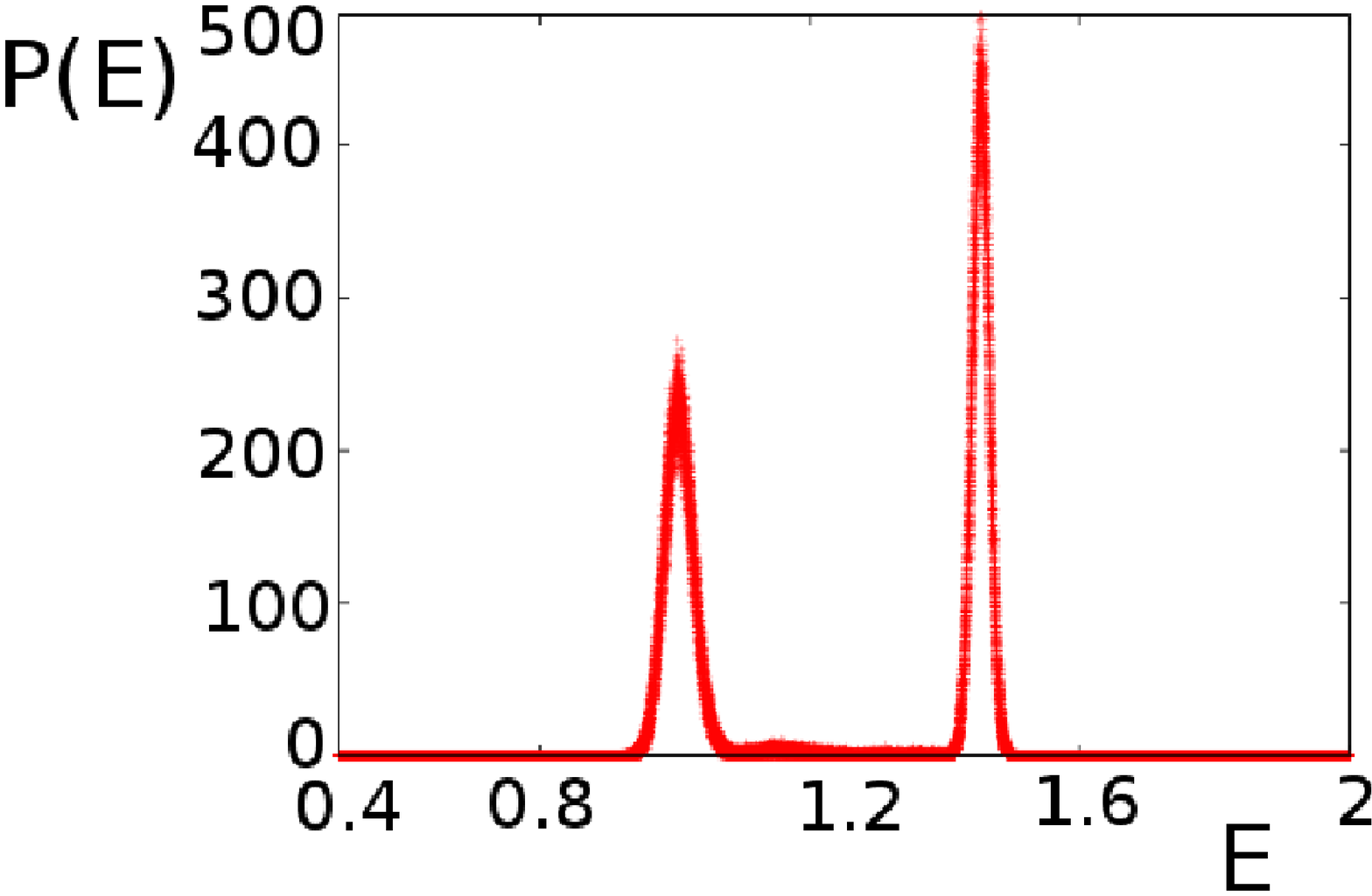}
\caption{\label{histd73}  (Color online) Energy histograms at $T_c=3.820$ (top) and  $T_c=2.694$ (bottom) for  $D=6$ (second-order transition)  and 7.3 (first-order transition), respectively.}
\end{figure}

Using the histogram technique \cite{Ferrenberg,Ferrenberg2,Ferrenberg3},  we explored the transition region to search for the nature of the transition.
For $D=6$, we obtain only a one-peak structure at the critical temperature (see Fig. \ref{histd73}, top). The energy histogram
taken at $T_c$ in the case $D=7.3$ exhibits a double-peak structure as shown in Fig. \ref{histd73} (bottom), confirming thus the first-order character of the transition \cite{Barber}.  

\begin{figure}
\centering
\includegraphics[width=6cm]{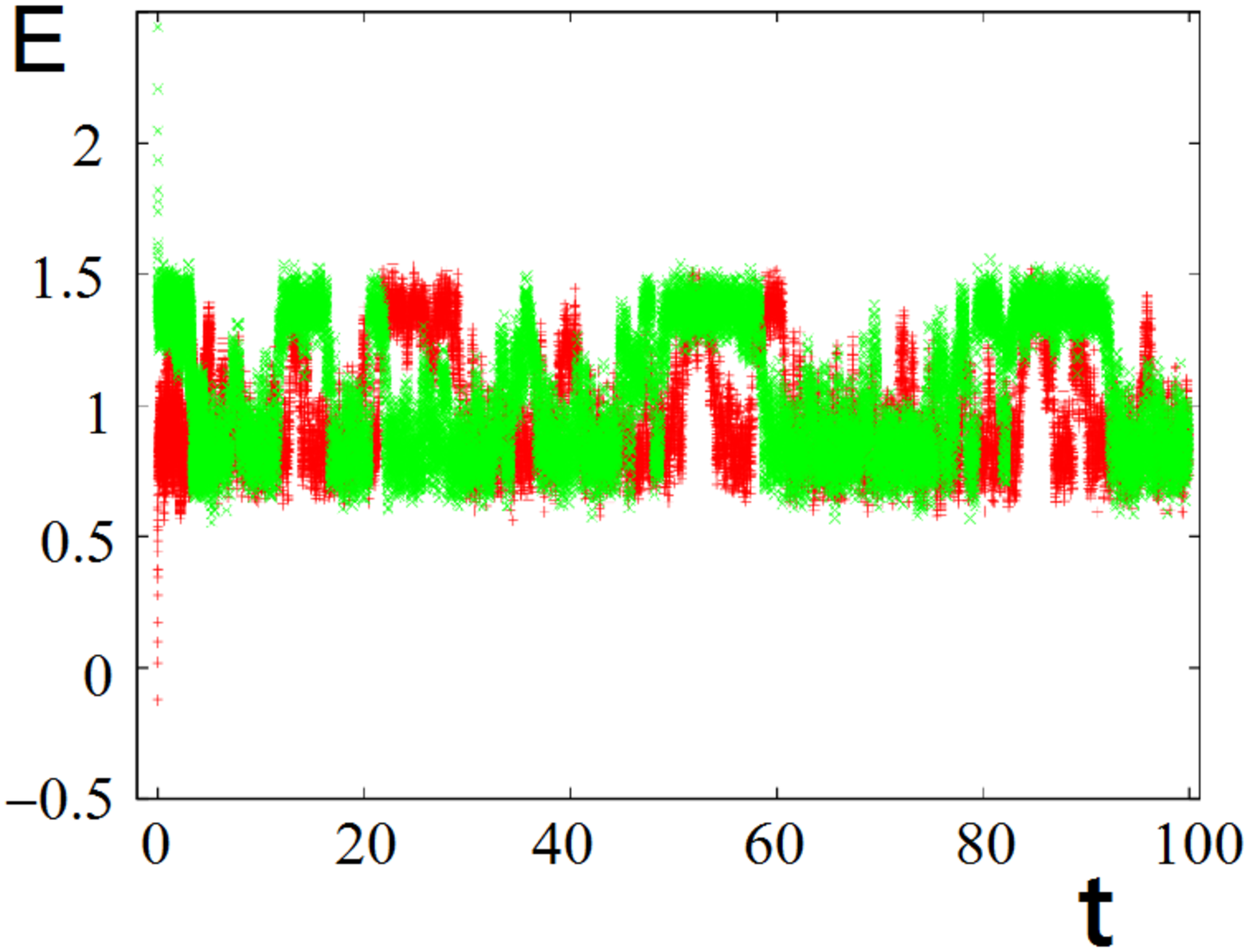}
\includegraphics[width=6cm]{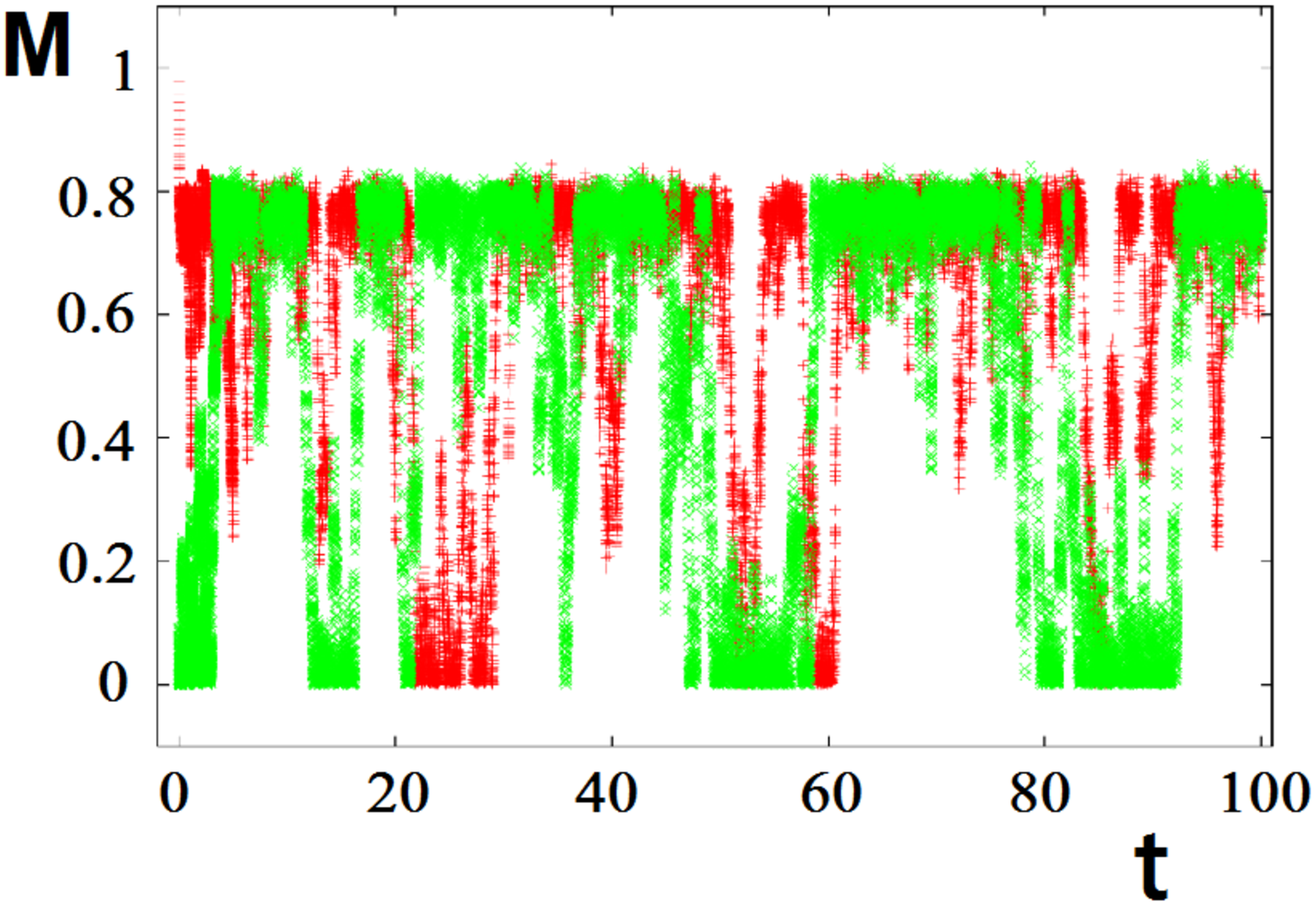}
\caption{\label{etime}  (Color online) Energy (top) and magnetization (bottom) versus MC time $t$ (in unit of $10^3$) at the transition temperature $T_c=2.694$ for $D=7.3$ (first-order transition). Note that the red and green curves are obtained with ferromagnetic and random initial configurations, respectively. See text for comments.}
\end{figure}

At a first-order transition the ordered and disordered phases coexist. In most cases, the system has mixed domains of two phases at the same time, the energy of the system is thus the average of the energies of the two phases $(E_1+E_2)/2$. It is however possible that at the transition the system goes back and forth between the two phases during the time evolution. This is what we observe here: we show in Fig. \ref{etime} how the energy and the magnetization evolve during the equilibrating time of $10^5$ MC steps/spin.
There are several remarks:

(i) In a general manner, in MC simulations a trick to use to check the equilibrium time is to do two simulations one with a random initial spin configuration and the other one with the ground-state configuration. We monitor  various physical quantities with time evolution. The equilibrium is attained when two initial spin configurations give the same results. We see in Fig. \ref{etime} that only after a few thousands of MC steps that the two initial configurations give statistically the same results.

(ii) The evolutions of $E$ and $M$ show bimodal distributions over periods of $\simeq 10^4$ MC steps. The time of $10^5$ MC steps for equilibrating and  $10^5$ MC steps  for averaging is thus sufficient as said above.

We have calculated the transition temperature with varying $D$ from 0 to 7.5.
The maximum value of $D$ for a 4-layer film is 7.5 above which there is no transition at all. This value depends on the film thickness. It comes from the fact that the maximum of $D$ should cancel the energy from $J$ and $K$ term. For example, with $L_z=4$, the energy of $J$ and $K$ terms is
\begin{eqnarray}
E_1&=&-(7J+7K)2\  (2\ \mbox{surfaces})\nonumber\\
&&-(8J+8K)2\  (2\  \mbox{interior layers})=-30\nonumber
\end{eqnarray}
where $J=K=1$. The energy from $D$ is $E_2=+2D$ (2 surface atoms)+$2D$ (two interior atoms)= 4$D$. The maximum of $D$ is determined by setting $E_1+E_2=0$, from which $D=30/4=7.5$.  The same calculation can be done for another thickness, yielding another value of maximal $D$.

\begin{figure}
\centering
\includegraphics[width=6cm]{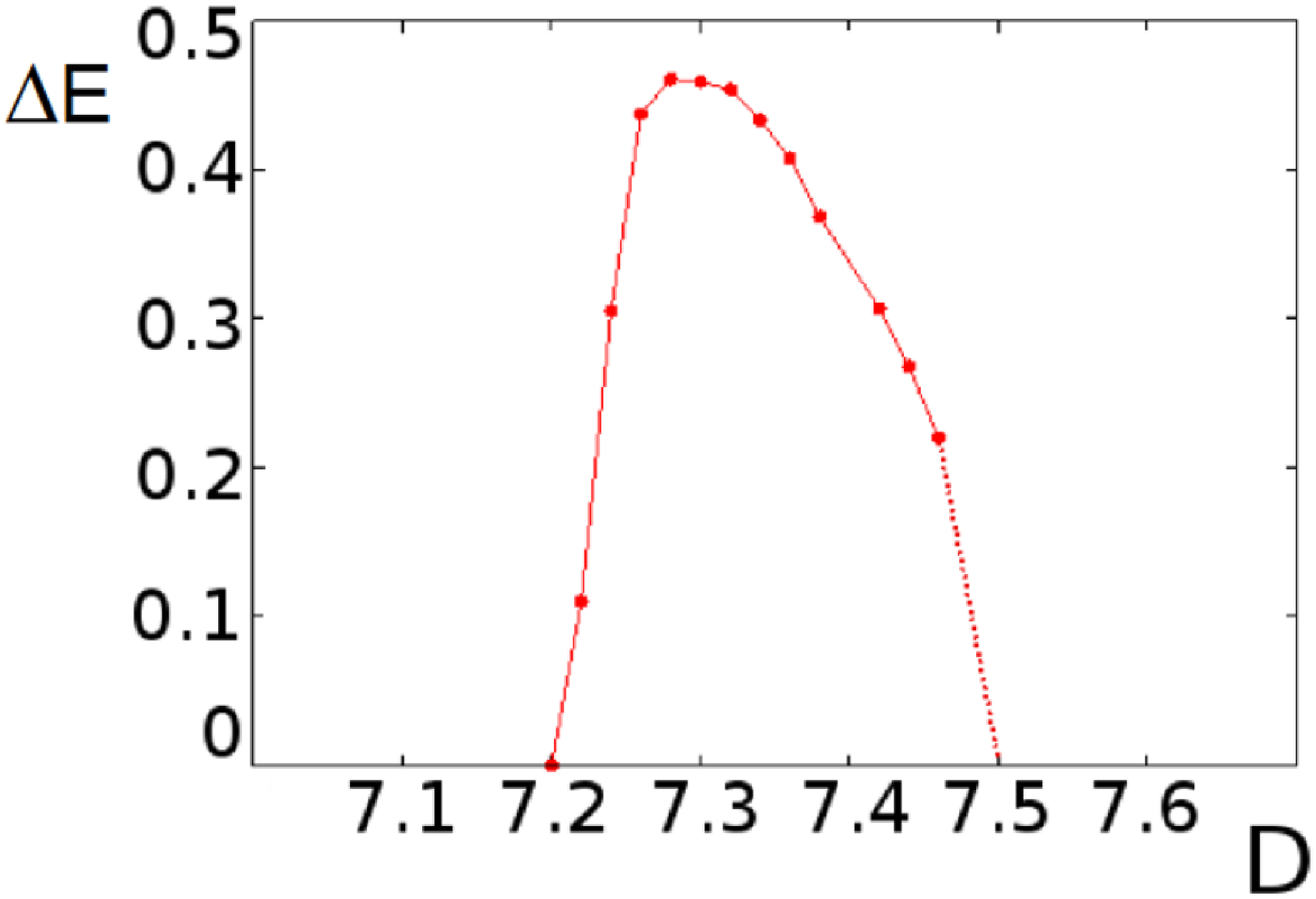}
\includegraphics[width=6cm]{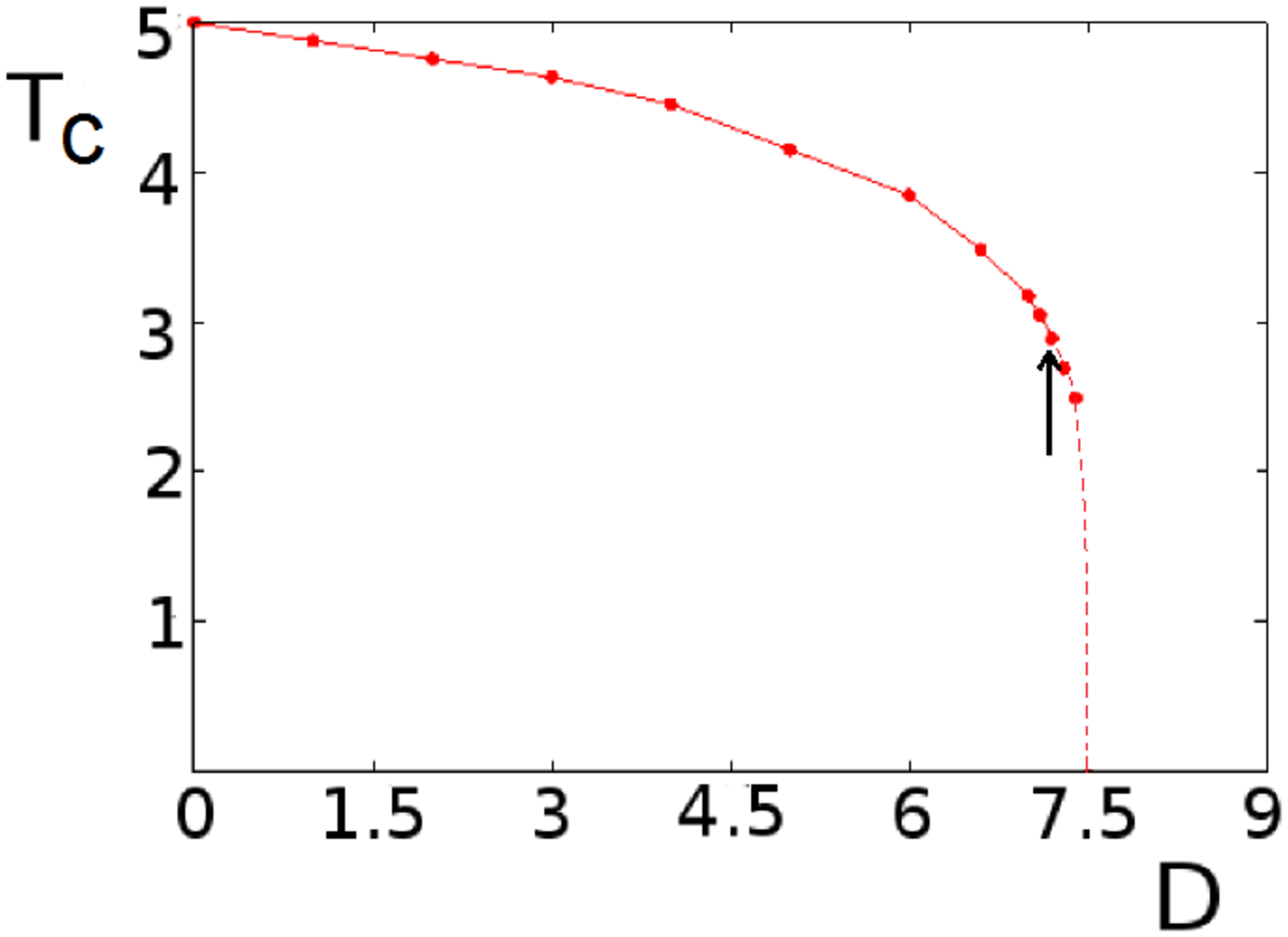}
\caption{\label{ftc}  (Color online) Latent heat $\bigtriangleup E$ (top) and $T_c$ (bottom) versus $D$, for $L_z=4$, $L=120$.  The critical value of D ($\simeq 7.2$) is shown by the vertical arrow. The dotted line between $D=7.4$ and 7.5 is extrapolated.}
\end{figure}

To determine the critical value of $D$, namely $D_c$, where the transition changes from second to first order, we follow the variation of the energy gap $\bigtriangleup E$ defined by the energy separation of the two peaks
in the energy distribution.  This gap is zero when the transition is of second order because the energy distribution is continuous. Using the histogram method with various values of $D$, we show  in Fig. \ref{ftc} (top)  the variation of $\bigtriangleup E$ versus $D$. As seen, $\bigtriangleup E$ is not zero for $D\in ]7.2,7.5[$. For $D\leq 7.2$ the   phase transition is continuous and for $D\geq 7.5$ there is no phase transition. We show in Fig. \ref{ftc} (bottom) $T_c$ versus $D$.

We note that the maximal value of $D$ and the tricritical value $D_c$ depend on the film thickness. We can notice that by looking at the bulk maximal value  $D=8$ ($L_z=\infty$) and $D_c=7.5$ as shown in Fig. \ref{bulk}. The four-layer film has $D_c=7.2$. So, when $L_z$ goes to infinity $D_c$ goes from 7.2 to 7.5.

\subsection{Size effect}
When the system size is infinite, in second-order phase transitions the correlation length is infinite at the
critical point. However, in first-order transitions the correlation length is finite at the transition temperature where the two phases coexist and the energy is discontinuous.  In simulations, in spite of the fact that we use periodic boundary conditions to mimic large systems, we cannot avoid finite-size effects on the results. The nature of the transition may not be detected at small system sizes.   It is therefore very important to measure the size effects in numerical simulations.
We have shown  in Fig. \ref{fed6} (bottom) the energy versus $T$ for $L=36$ and $L=300$. The size effect is extremely small. The transition remains continuous though one observes a change in the slope of the curve which is steeper for the larger size.  In the first-order region, the energy and magnetization are already discontinuous even for  $L$ as small as 36.

\begin{figure}
\centering
\includegraphics[width=6cm]{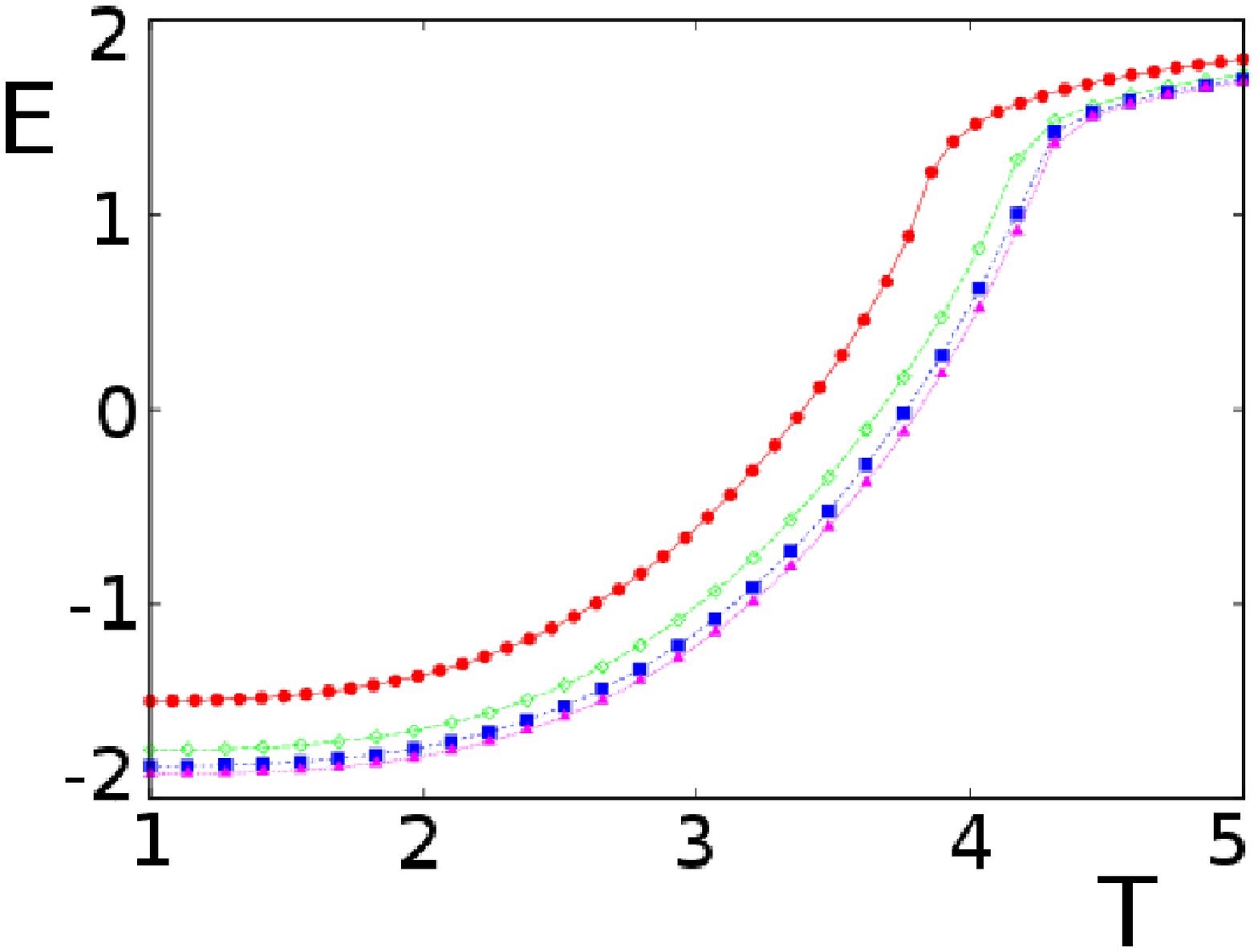}
\includegraphics[width=6cm]{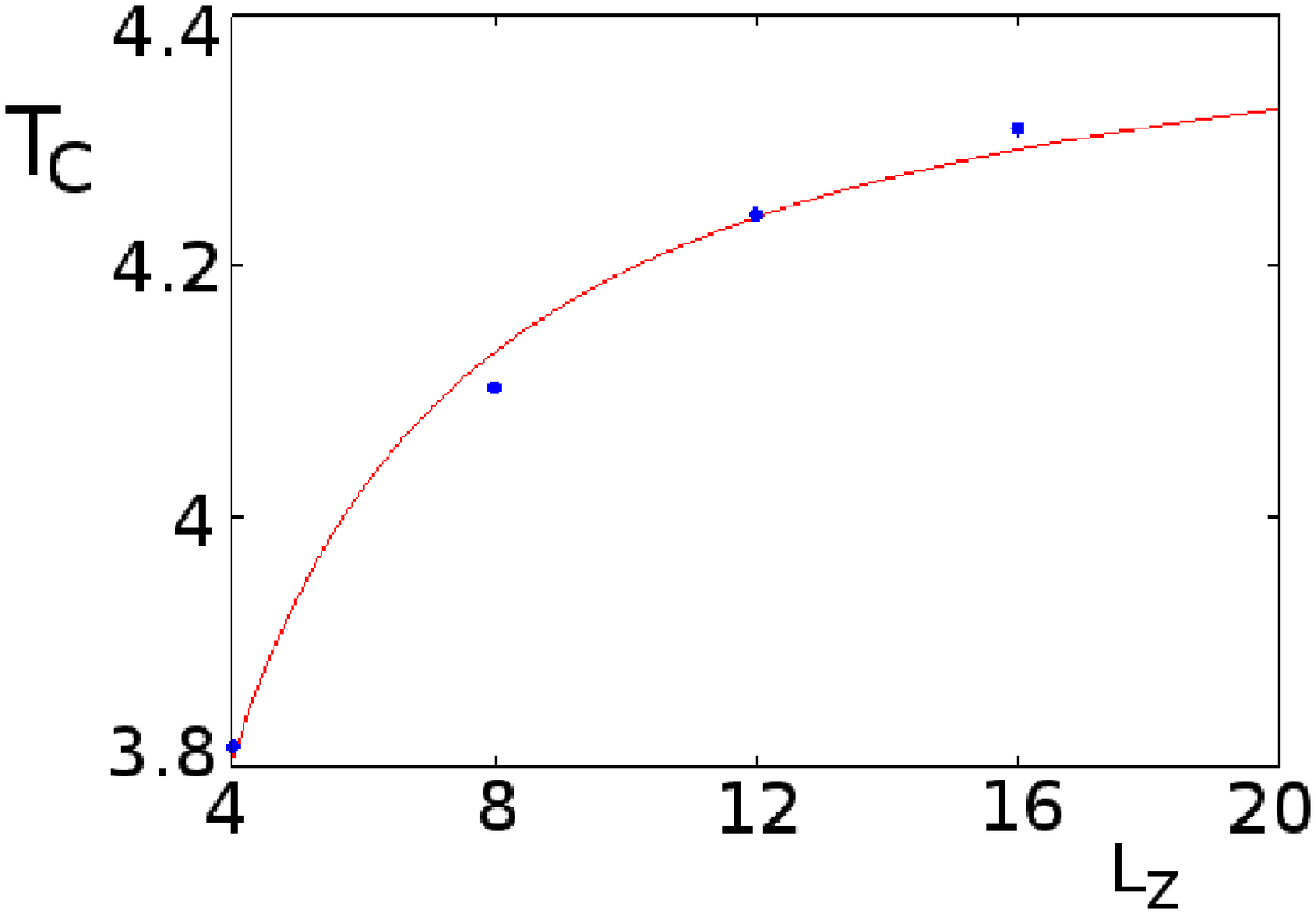}
\caption{\label{fsef}  (Color online) Top: Energy versus $T$ for $L_z$=4 (red), 8 (green), 12 (blue) and 16 (magenta) (from above), with $L=120$, and $D=6$. Bottom: $T_c$ (blue points) versus $L_z$ for $D=6$. The continuous red line is the least mean-square fit. }
\end{figure}

The film thickness affects, on the other hand, the value of the transition temperature $T_c$ as seen in Fig. \ref{fsef}.  As $L_z$ increases, the transition temperature tends to that of the bulk.  We have used the least mean-square fit with the form:
\begin{equation}\label{TcLz}
T_c(L_z)=T_c(\infty)-\frac{A}{L_z}
\end{equation}
where $A=2.692\pm 0.165$ and $T_c(\infty)=4.455\pm 0.024$.

The way how $T_c$ increases with increasing thickness is characterized by constant $A$ in the above equation of $T_c(L_z)$. This constant is different from one material to another depending on the coupling between film layers.  In some materials $A$ is very small, meaning that inter-layer coupling is very small. This is not the case here in spite of the fact that there are only two nearest neighbors for each interior atom on the $z$ axis (with only one for a surface atom).  Knowing how $T_c$ varies with the film thickness can help determine the inter-layer coupling.  

At this stage, let us discuss about the criticality of the transition in the second-order region. If we compare Eq. (\ref{TcLz} with the finite-size scaling relation
\begin{equation}
T_c(L)=T_c(\infty)+AL^{-1/\nu}
\end{equation}
we see that $\nu=1$ which is the 2D Ising universality exponent. This is in agreement with Ref. \onlinecite{Pham1}: when the film thickness becomes small the critical exponents tend to the 2D criticality.

\subsection{Surface effect}
\begin{figure}
\centering
\includegraphics[width=6cm]{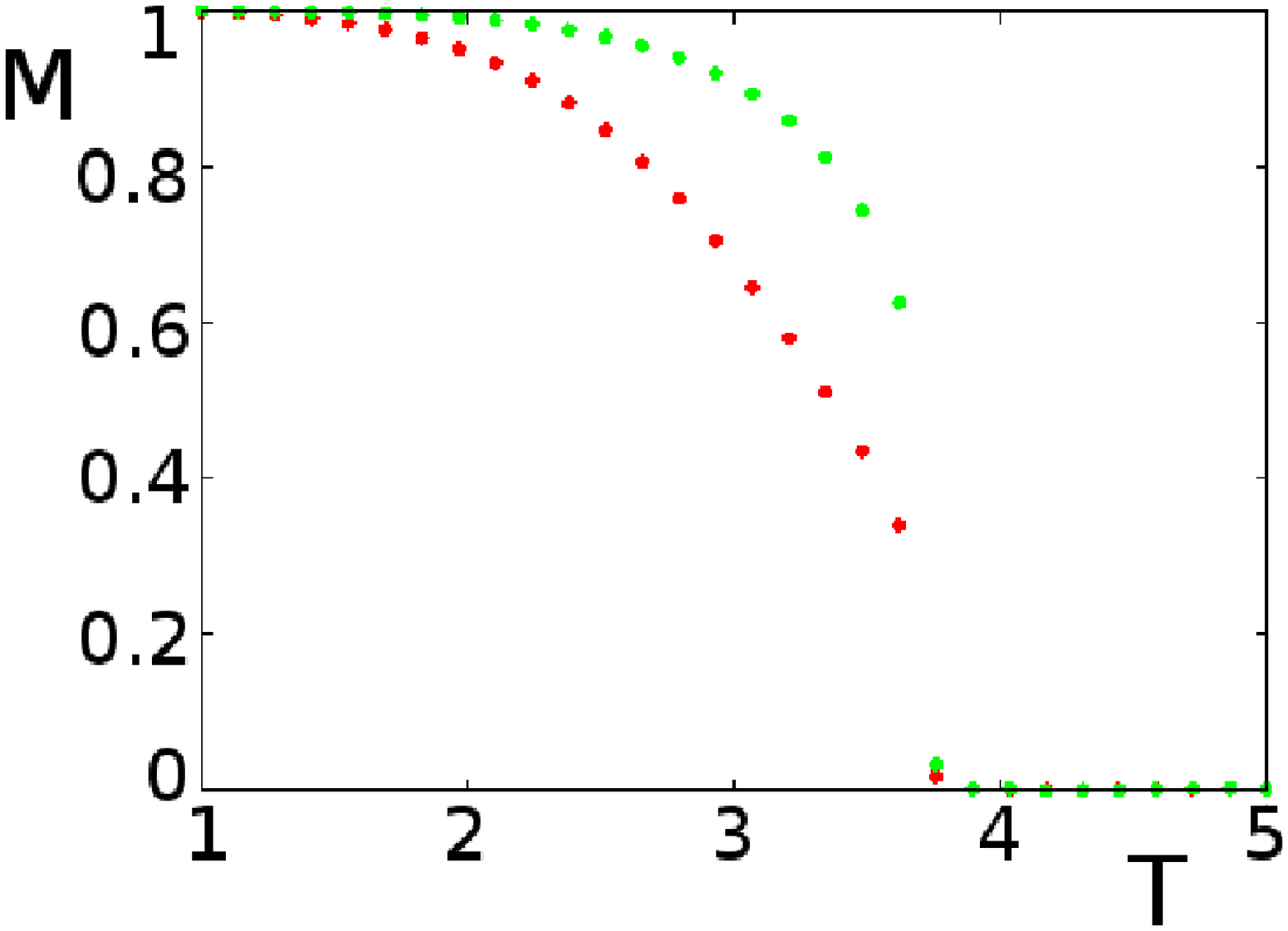}
\includegraphics[width=6cm]{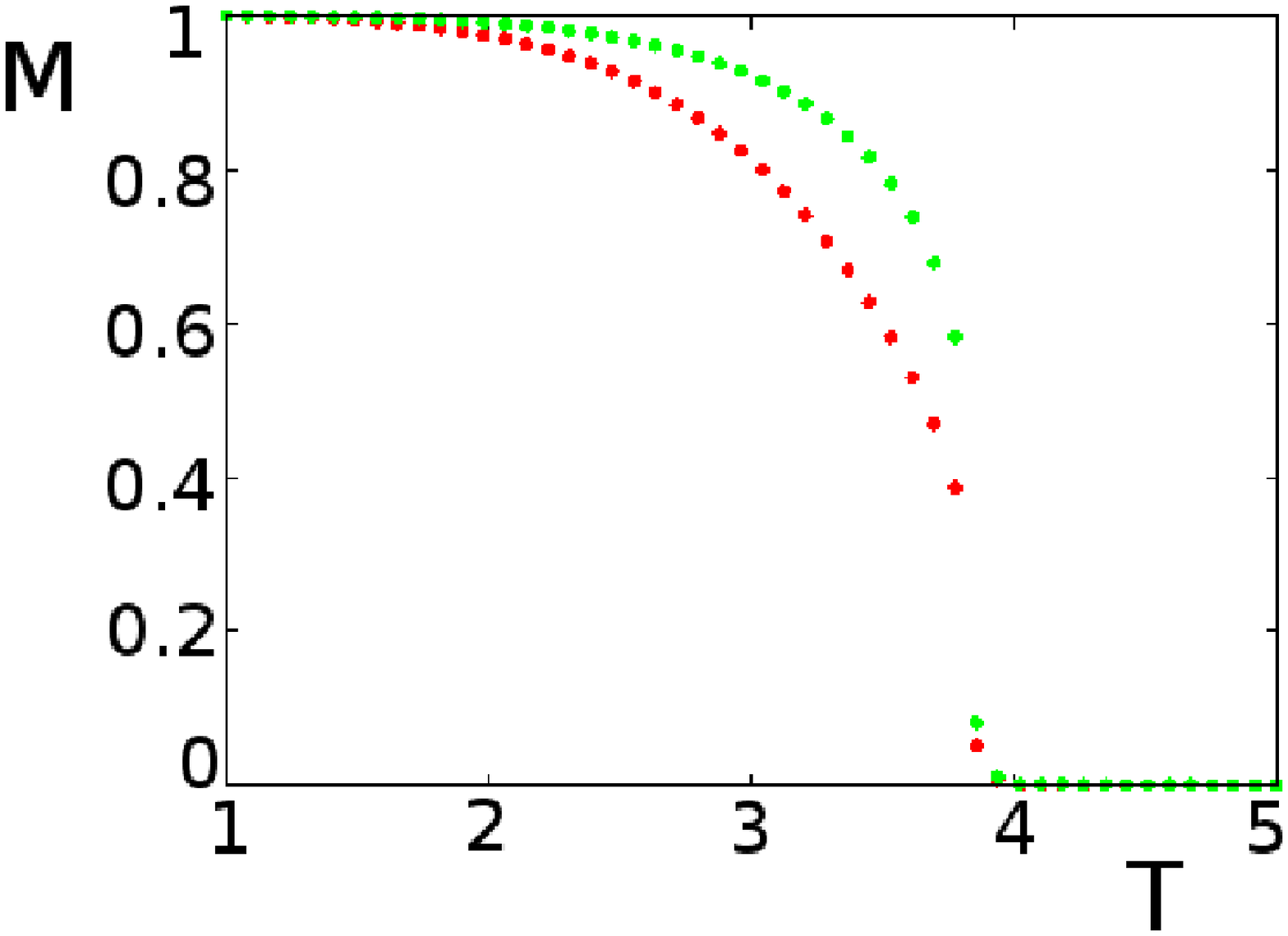}
\includegraphics[width=6cm]{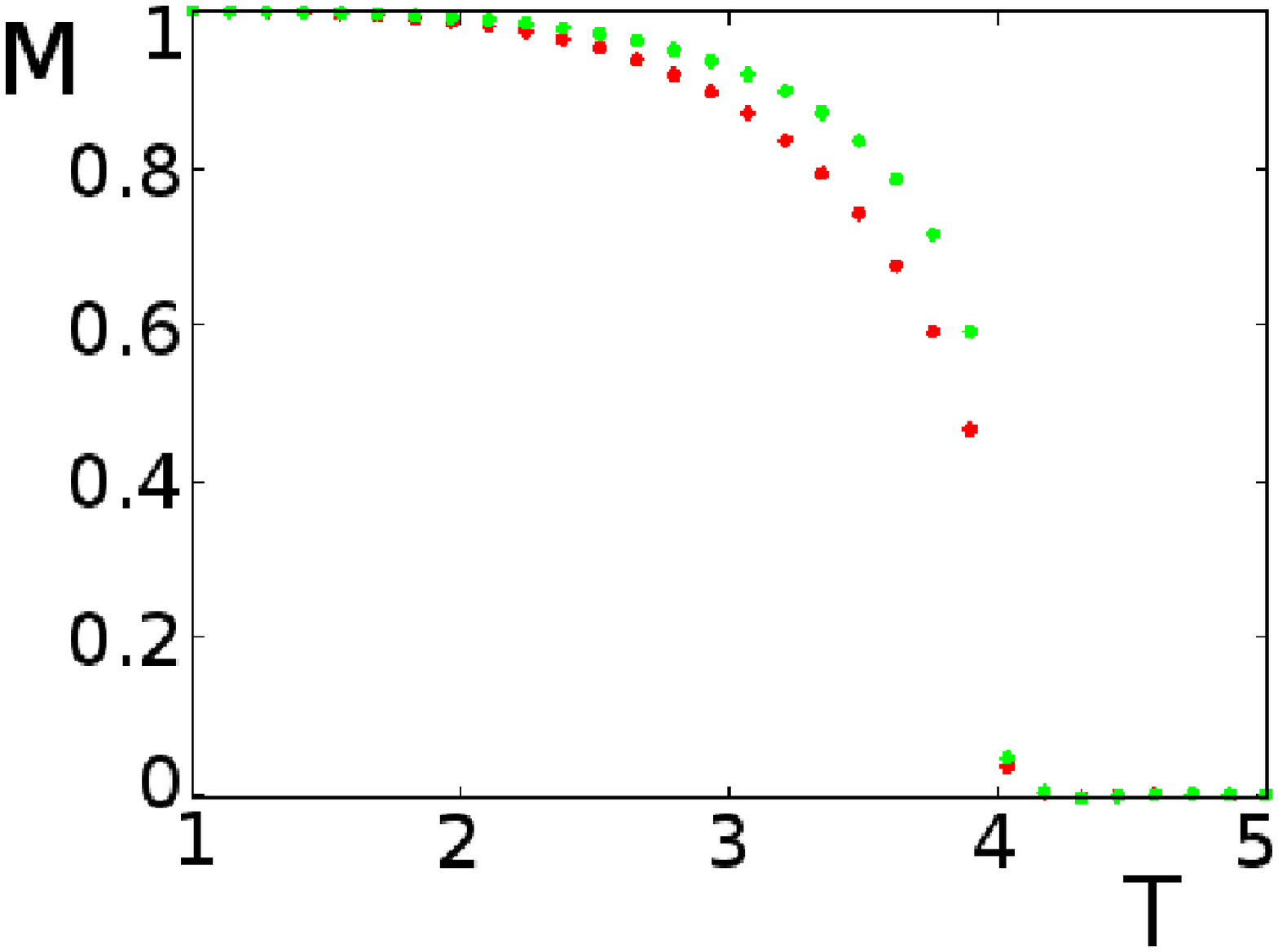}
\includegraphics[width=5.5cm]{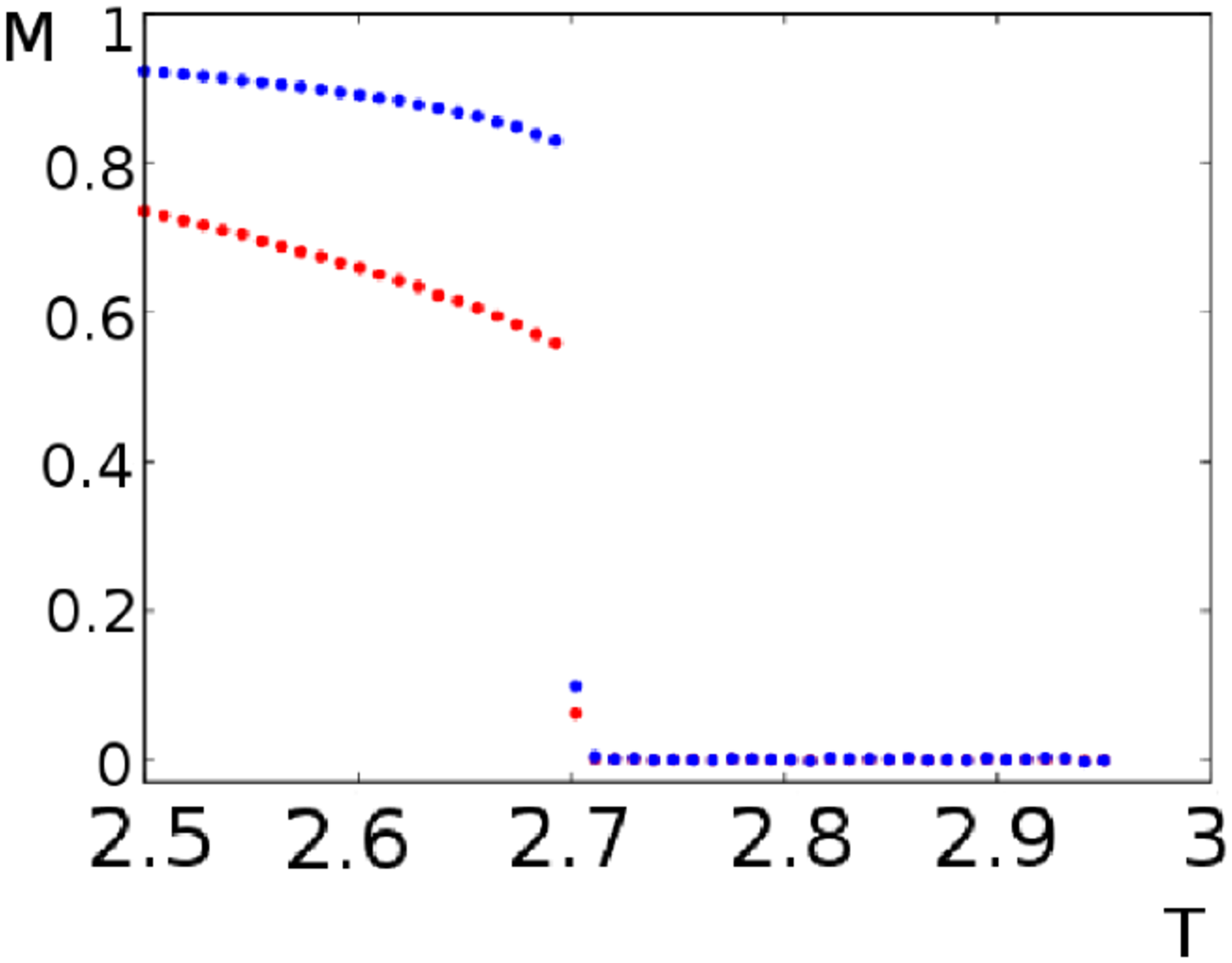}
\caption{\label{layerM}  (Color online) Layer magnetization of the first (red) and second (green) layers versus $T$. The first three figures (from top) are for: $\alpha_s=0.8$, 1, and 1.2, with $D=6$. The bottom figure represents a first-order case where $\alpha_s=1$ and $D=7.3$: red (blue) symbols indicate the first (second)-layer magnetization.}
\end{figure}

So far, we have supposed $J=K$ for any NN spin pair in the film. We investigate now the surface effect due to the surface parameter $K_s$ taken to be different from $K$.
We write the biquadratic surface and bulk parts as follows:
\begin{equation}
 \alpha_b\sum\limits_{i,j} S_i^2S_j^2+\alpha_s\sum\limits_{i',j'} S_{i'}^2S_{j'}^2
\end{equation}
where $\alpha_b=K/J$ and $\alpha_s=K_s/J$ denote respectively the bulk and surface interactions and $\sum\limits_{i',j'} $ denotes the sum over NN spin pairs in the surface layer. We take $\alpha_b=1$.
Let us show the magnetization of the first and second layers in  Fig. \ref{layerM} for several values of $\alpha_s$.  As seen, the weaker the surface interaction is, the smaller the surface magnetization becomes.  Only when $\alpha_s$ is much larger than 1,  the surface-layer magnetization becomes larger than the second-layer (not shown).
For the first-order region, the surface and interior magnetizations have discontinuities at the transition as expected.

\begin{figure}
\centering
\includegraphics[width=6cm]{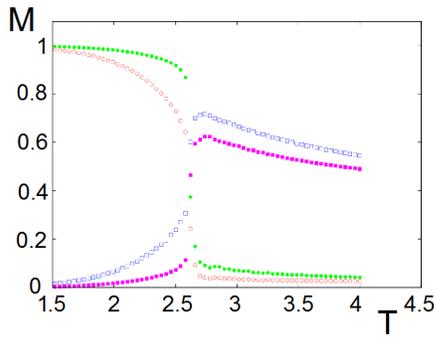}
\caption{\label{figonezero}  (Color online) The normalized numbers of spins $\pm 1$ (He-4) and spins zero (He-3) versus $T$ for the first and second layers. Red void circles and green circles represent the number of He-4 (spins $\pm 1$) on the first and second layers, while blue void squares and magenta squares represent the number of He-3 (spins zero) on the first and second layers, respectively.  See text for comments.}
\end{figure}

We show now the average number of spins $\pm 1$ and the average number of spins zero in each layer versus $T$ in Fig. \ref{figonezero} at the first-order transition with $D=7.3$.
They are defined as $M_{1,2}(\pm1)=\langle\sum_{i}S_i [\delta (S_i,1)+\delta (S_i,-1)]\rangle/L^2$ where the sum is made for each layer: $M_1(\pm1)$ [$M_2(\pm1)$] corresponds to the surface (second) layer. For spins zero,   $M_{1,2}(0)=\langle\sum_{i} \delta (S_i,0)\rangle/L^2$. Several remarks are in order:

(i) Below the transition temperature, the ordering results from spins $\pm 1$. The number of spins zero increases slowly from 0 at $T<T_c$ but becomes dominant for $T>T_c$.

(ii) At $T<T_c$ the surface has a smaller number of spins $\pm 1$ than the second layer, namely there is a deficit of He-4 at the surface. While, the number of spins zero is larger at the surface than in the second layer.

\section{Conclusion}
We have investigated in this paper the BEG model used for a thin film of stacked triangular lattices with a thickness $L_z$. There are three important aspects of our results for thin films:

(i) the nature of the first-order phase transition in a region of the phase space is conserved down to a 4-layer film, unlike in other systems where bulk first-order transition becomes second-order with small thickness  \cite{Pham2},

(ii) the cross-over from second-order to first-order transition in the bulk is also conserved in thin films as shown above. The anisotropy of the BEG Hamiltonian affects the nature of the phase transition as it has been observed in the bulk case of simple cubic lattice \cite{Puha}: in 4-layer triangular films, for $D\leq 7.2$ the transition is continuous and for $7.2<D<7.5$ the transition is of first order. This has been confirmed with the histogram technique where the latent heat can be measured with precision,

(iii) The surface effect on the layer magnetizations has been shown.  The surface magnetization is smaller than the interior layer if the surface interaction is not so large. If we map the BEG model into a mixing of He-3 and He-4, then near the surface there is a He-3 enrichment (normal liquid) in a film at low temperatures.  This point is new with respect to the bulk properties where the mixing of two liquids is uniform over the system.

\acknowledgments
SEH acknowledges  a financial support from Agence Universitaire de la Francophonie (AUF).

\end{document}